# Energy Reconstruction in Analysis of Cherenkov Telescopes Images in TAIGA Experiment Using Deep Learning Methods

E. O. Gres[a,b,∗] and A. P. Kryukov[b]

[a] *Applied Physics Institute of Irkutsk State University,*
  *20, Gagarin boulevard, Irkutsk, 664003, Russian Federation*

[b] *Skobeltsyn Institute of Nuclear Physics, Lomonosov Moscow State University,*
  *1(2), Leninskie gory, Moscow, 119991, Russian Federation*

  *E-mail:* greseo@mail.ru, kryukov@theory.sinp.msu.ru

Imaging Atmospheric Cherenkov Telescopes (IACT) of TAIGA astrophysical complex allow to observe high energy gamma radiation helping to study many astrophysical objects and processes. TAIGA-IACT enables us to select gamma quanta from the total cosmic radiation flux and recover their primary parameters, such as energy and direction of arrival. The traditional method of processing the resulting images is an image parameterization - so-called the Hillas parameters method. At the present time Machine Learning methods, in particular Deep Learning methods have become actively used for IACT image processing. This paper presents the analysis of simulated Monte Carlo images by several Deep Learning methods for a single telescope (mono-mode) and multiple IACT telescopes (stereo-mode). The estimation of the quality of energy reconstruction was carried out and their energy spectra were analyzed using several types of neural networks. Using the developed methods the obtained results were also compared with the results obtained by traditional methods based on the Hillas parameters.

[∗]Speaker

## 1. Introduction

Gamma-ray astronomy is one of the most important area of observational astrophysics. The study of processes occurring in such physical systems and processes as stellar explosions, powerful high-velocity outflows forming in the vicinity of super massive black holes of active galaxies and more are important for creating a complete picture of the Universe evolution [1]. In addition, the list of tasks of modern gamma astronomy includes the experimental verification of hypotheses of fundamental physics (for example, hypotheses about the nature of dark matter, quantum gravity, etc.). Ultra high energy gamma radiation (tens and hundreds of TeV) is a unique source of information about the processes that occur in such astrophysical objects.

Physicists develop new methods, create various installations and conduct experiments to observe gamma radiation and measure its parameters (energy, direction of the arrival and more). At present it's possible to register gamma-ray photons of ultra high energies only from the Earth's surface. The main instruments for observing gamma rays are Imaging Atmospheric Cherenkov Telescopes (IACT). At the moment there are several experiments (MAGIC, H.E.S.S., VERITAS, CTA, TAIGA) that register gamma using IACT and analyze images using traditional methods (the so-called Hillas parameter method) [2] – by parameterizing images and imposing conditions on these parameters. Currently, the use of machine learning has also been developed actively in image processing in gamma astronomy, which allows to analyze large amounts of data with improved results compared to traditional methods. For example, in MAGIC and H.E.S.S. the application of machine learning methods has shown promising results in the problem of classification of registered by telescopes particles (i.e. selection of gammas and hadrons) [3-6].

The objective of this work is to develop and study deep machine learning methods for processing and analyzing TAIGA-IACT data. TAIGA-IACT is a part of the hybrid installation TAIGA (Tunka Advanced Instrument for cosmic ray physics and Gamma ray Astronomy) located in the Tunka valley of the republic Buryatia near Lake Baikal [7]. These telescopes have large spherical segmented mirrors with a camera consisting of photomultipliers (PMTs) in the focus of the mirrors. The camera is a matrix of almost 600 PMTs. The main processing task of the TAIGA-IACT is to separate gamma events from the cosmic ray background and reconstruct the parameters of the primary particle. Currently 3 telescopes have been installed and are in operation.

It's worth noting that research on the application of machine learning methods for this installation was previously carried out in this works [8-12]. In these works neural networks of the same configuration were considered or the problem of classifying gamma events from hadrons was mainly studied. And in general, they aimed to demonstrate the possibility of applying these methods to solve problems of gamma-ray astronomy at a qualitative level. In this work the analysis and energy reconstruction of Monte Carlo data will be carried out by several deep learning methods in the case of event registration by one (mono mode) and several TAIGA-IACTs (stereo mode).

## 2. Model data

In the work a study was conducted on energy reconstruction based on labeled model data generated by CORSIKA [13]. The description of the datasets used and their division into training and test sets are presented in Table 1. For the mono mode, the problem of energy reconstruction was considered in the case of set of mixed events (when events from gammas and protons are presented in set) and in the case of a set of gammas only. In the case of stereo mode, only gamma events were considered. Studying on gamma sets is due to the fact that an attempt to recover energy for a mixed set gives large errors (this is outlined below), and also that before reconstructing energy all recorded events are classified with subsequent selection of gammas.

Table 1. Description of model samples used in training and testing various deep learning methods under different observation modes

| Mode | Total events (gammas/protons) | Train/test separation | Energies, TeV |
|---|---|---|---|
| Mono | 200 000 (100 000 / 100 000) | 160 000 / 40 000 | Protons: 5-100 Gammas: 2-50 |
| Stereo | 30 000 – «mono» 14 800 – «stereo-2» 7 700 – «stereo-3» (only gammas) | Separation 3:1 in each case | 1-50 |

As can be seen from the table, the training sample of the stereo mode has quite a few thousands events, so it was expanded by mixing the input images. For example, in the case of two telescopes, one event has two images, one of which belongs to the first telescope, the second - to the second IACT. When shuffled, it was assumed that the first image is now the image of the second telescope and vice versa. This approach is valid in the case of telescopes that do not differ much from each other in their structure and internal event triggers. In our case these conditions are satisfied.

The images were also pre-processed, which included cleaning, pixelation and logarithmic scaling. When cleaning, the noise pixels in the image were reset to zero. The cut-off threshold of cleaning was 3 photoelectrons. The need for pixelation is due to the fact that deep learning methods used for image processing (so called convolutional neural networks) are not able to work with hexagonal image structure. The hexagonal image structure is a consequence of the use of PMTs in the IACT camera. There are a lot of way of transition to rectangular images, in our work we apply axial coordinate transformation [14, 15]. With logarithmic scaling the amplitudes of pixels $x_i$ were transformed as follows:

$$\tilde{x}_i = \frac{1}{K} \ln(1 + x_i) \quad (1)$$

where $i$ is a number of pixel in the image, $\tilde{x}_i$ is scaled pixel amplitude. Normalization constant $K$ is associated with the maximum possible value of the pixel amplitude and brings the amplitude into the range of values from 0 to 1 to improve the training of neural networks. For our case $K$ is equal to 9.

## 3. Deep learning methods

Machine learning is a scientific discipline that uses a sample–based learning method instead of explicitly programming a computer system. Deep learning implies machine learning using neural networks (NN) – multilayer graphs, where a mathematical model of a neuron is located at the node of each graph [16, 17].

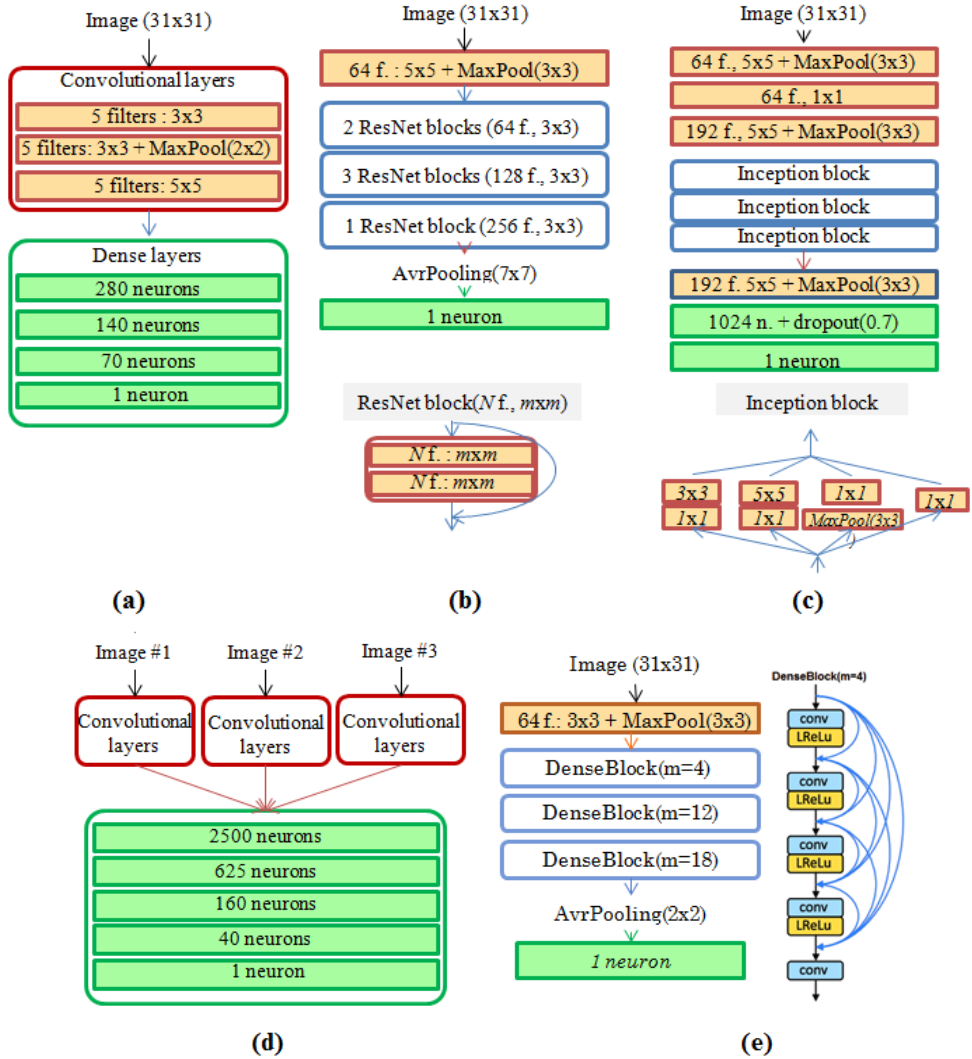

**Figure 1:** Schematic architectures of convolutional neural networks applied in the work for data analyses of mono-mode (a-c) and stereo-mode (a,d-e): a) User linear CNN; b) Simplified ResNet; c) Simplified GoogLeNet; d) Multichannel User CNN; e) Simplified DenseNet

To process telescope images convolutional neural networks (CNNs) [17, 18] were used, the structures of which for the cases of mono- and stereo-mode are shown in Figure 1. The architectures were written in Python using TensorFlow and Keras libraries [18, 19]. The initialization of the weights is the uniform initialization of Xavier, the activation function between layers for User models is ReLU. On the output layer, the activation function is linear, the error function was defined as the mean squared error. The figure 1 shows that along with User CNN ResNet [20], GoogLeNet [21] and DenseNet [22] structures were considered. These architectures have shown some of the best results in image classification [22, 23] and they use various methods of layers connection to stabilize and improve training. However, for an adequate comparison with the User model, their structures were simplified so that the number of weight coefficients for each network was approximately 2 million.

GoogLeNet and ResNet structures were used to analyze mono-mode data, the linear User model was modified for stereo mode (see Fig. 1, d) by adding additional input channels. Multichannel User model was used in stereo mode processing together with the linear User CNN and DenseNet. For linear models with a single input, images from several telescopes were overlaid on each other with coinciding the center of the cameras and summing the pixel amplitudes. The overlay occurred after the images were cleaned, after which they were pixelated and scaled by (1).

## 4. Energy reconstruction results

The energy reconstruction by the above-described CNNs structures was considered in two cases of event registration: mono mode and stereo mode. To indicate how many telescopes registered the event at the same time, the «stereo-N» mode will be abbreviated, where $N$ indicates the number of triggered telescopes. For estimation of the energy reconstruction quality of each event and the energy spectrum as a whole, the relative error $Rel\_err$ and criterion $\chi^2$ were used, respectively:

$$Rel\_err = \frac{|E_{pred} - E_{true}|}{E_{true}}, \quad (2)$$

$$\chi^2 = \sum_{i=1}^{k} \frac{((c_{rec})_i - (c_{MC})_i)^2}{(c_{MC})_i} \quad (3)$$

where $E_{pred}$ is an energy predicted by CNN, $E_{true}$ is true energy value of event, $k$ is the number of bins in the spectrum histogram, $c_{rec}$ is the number of events in bins in the case of the reconstructed spectrum, $c_{MC}$ is the number of events in bins of the model spectrum.

### 4.1 Mono mode

When considering the issue of energy recovery for the set with mixed events, a linear User CNN was trained and tested. The median relative error is 31% in this case. When energy reconstructed only for gamma events, the quality of energy reconstruction is noticeably improved (Fig. 2): the median relative error is reduced to 22-26% (the best result is achieved with the GoogLeNet). Strong deviations are observed at the edges of the spectrum associated with the approximating function of neural networks when reconstructing the spectrum. The criterion $\chi^2$ reaches values of the order of several thousand. However, results demonstrated that a good matching of the spectra is observed in a narrow energy range.

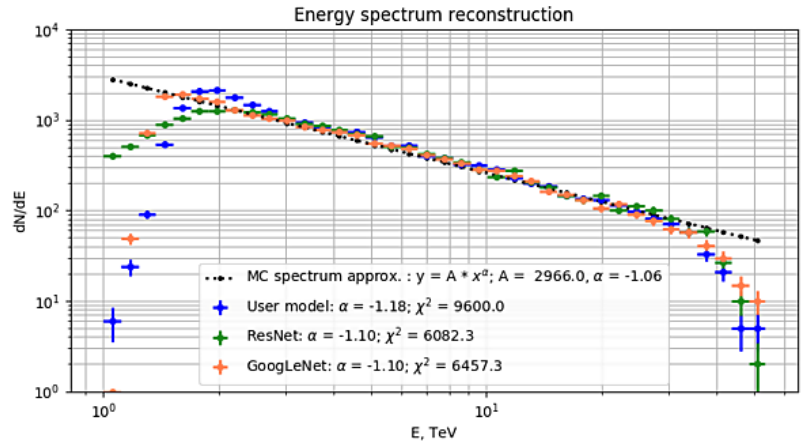

(a)

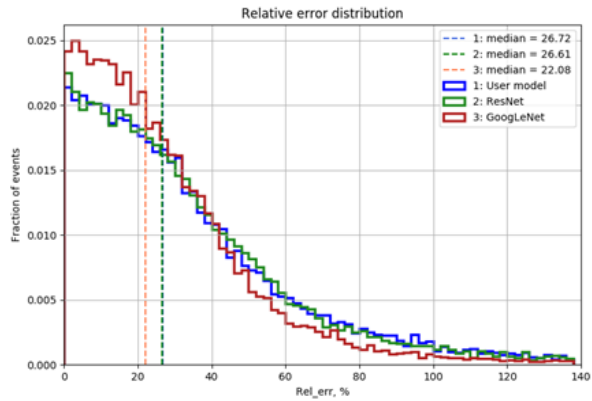

(b)

**Figure 2**: Energy spectrum reconstruction of gamma events (a) and relative error distribution (b) with different CNNs in case monoscopic observations of TAIGA-IACT

### 4.2 Stereo mode

When studying the quality of energy recovery in mono and stereo modes, a Multichannel User CNN was applied. The results of the spectrum reconstruction and the distribution of relative errors are shown in Figure 3. The estimation showed that the $\chi^2$ values in mono mode are 1 546, in the case of «stereo-2» – 495, in «stereo-3» – 156. The relative error decreased from 26% to 15%.

We also wanted to see if it was possible to improve the reconstruction accuracy in the stereo mode with other CNN models, therefore, further we considered only the «stereo-3» mode and the previously mentioned linear User CNN and DenseNet (see Fig. 1, a and Fig. 1, e). To compare the models we had to modify the multichannel CNN by adding three dense layers in the input channels before they were connected (the number of neurons: 440, 170 and 70) and reducing the number of neurons in dense layers after the connection (their number became 100, 40 and 1). Thus, the number of weight coefficients in multichannel CNN decreased by 10 times.

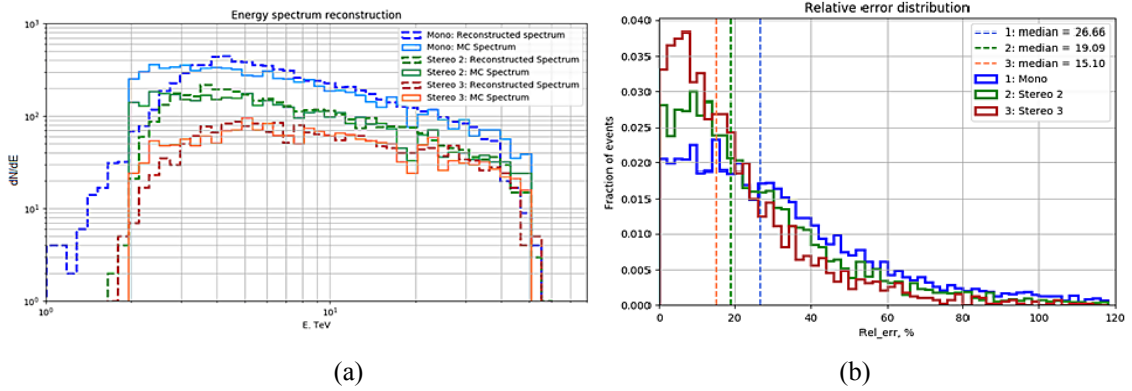

(a)         (b)

**Figure 3:** Energy spectrum reconstruction of gamma events (a) and relative error distribution (b) with multichannel user CNNs in case stereoscopic observations of TAIGA-IACTs

Reconstruction results are shown in Figure 4. It can be seen that each of the CNNs gives small differences in the reconstruction of the spectrum shape, while in determining the relative errors in the reconstruction of each event DenseNet significantly reduces the error: it has become 12%. The form of data submission gives a slight improvement in the result of user networks: an improvement in spectrum reconstruction is seen (the $\chi^2$ criterion has decreased from 156 to 87), but the error doesn't decreased (also 15%).

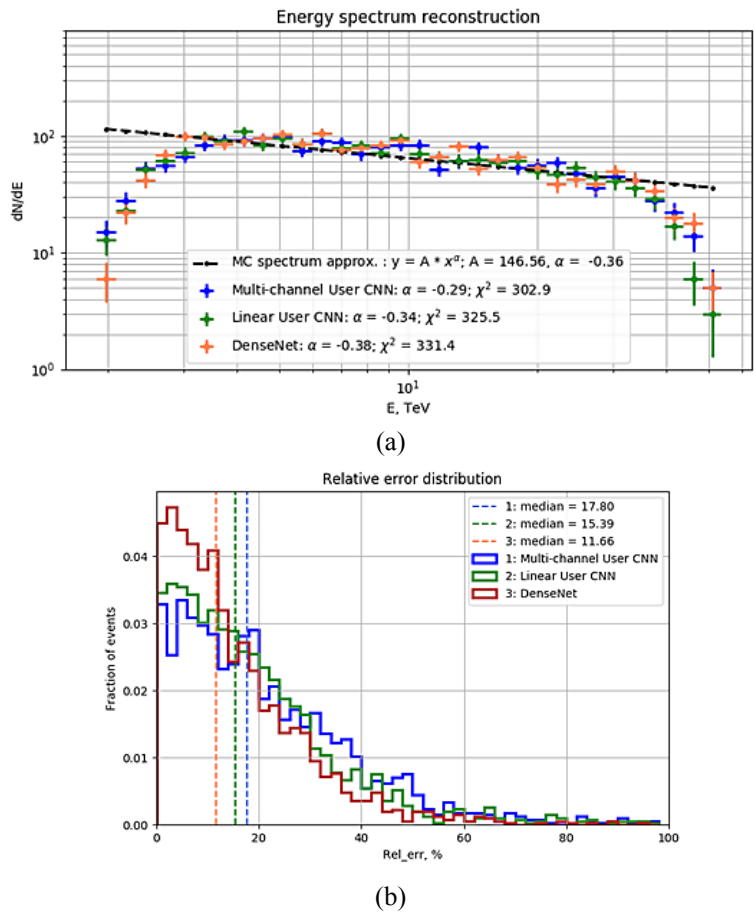

(a)

(b)

**Figure 4:** Energy spectrum reconstruction (a) and relative error distribution (b) with different CNNs in case stereoscopic observations with three TAIGA-IACTs («stereo-3»-mode)

## 5. Comparison of deep learning method and traditional energy reconstruction method

To estimate the possibility of applying the deep learning methods to experimental data results were compared with the traditional reconstruction method. The dependence of the event energy on some Hillas parameters, such as image brightness, the distance of the spot center of gravity from the camera center, and also on some extensive air shower (EAS) characteristics (the height of the EAS maximum, etc.) is searched in the traditional method of energy reconstruction by analyzing Monte Carlo data for each telescope [24]. After tables of correspondence between Hillas parameters and energy are compiled.

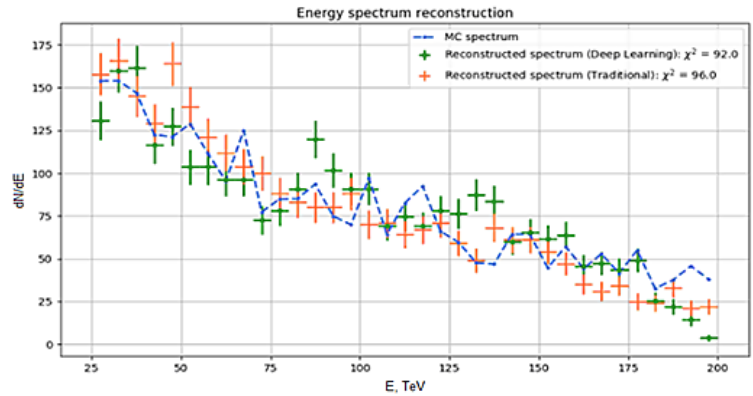

(a)

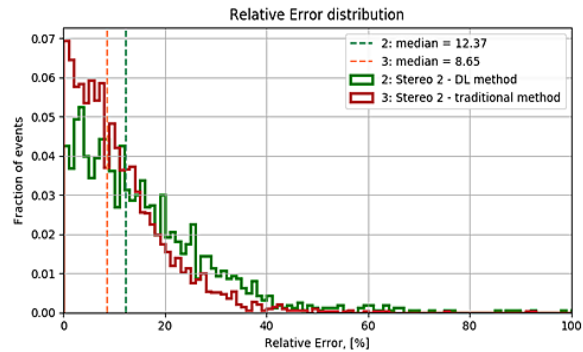

(b)

**Figure 5**: Comparison of energy spectrum reconstruction (a) and relative error distribution (b) using deep learning (Multichannel User CNN, red color) and traditional reconstruction methods (green color)

A new simulated dataset was obtained: this set included joint events from five telescopes (approximately 5 thousand events per pair), only a pair of IACT#1 and IACT#2 were processed with traditional method [25]. The energy range for these events ranged from 25 to 200 TeV. 1/3 of the events of the IACT#1 and IACT#2 pair were taken as a test set. However, to expand the training sample, events were taken from other pairs of telescopes that were at the same distance from each other (approximately 320 m). Thus, the training and test sample consisted of 16 620 and 1 598 events, respectively.

A two-channel User CNN was selected for training (see Fig. 1, d, instead of three inputs there are two of them). The regression results are shown in Figure 5. From the quantitative evaluation of the spectra it can be seen that CNN result demonstrate a good agreement with the results of the standard method, but the median error in determining the event is slightly worse: 13% compared to 9%. It's worth noting that the User model was used, so it's possible that when using the above structures and another data submission (overlaying images of several telescopes on one), the relative errors will be approximately the same as with standard methods.

## 6. Conclusion

Deep learning methods for processing and analysis the TAIGA-IACT data was applied in this work. Along with the usual linear architectures of convolution neural networks, which were considered in previous works on the application of machine learning to the images of this installation, well-known ResNet, GoogLeNet and DenseNet structures were programmed, trained and tested to solve the problem of energy reconstruction in the processing and analysis the TAIGA-IACT data. Also for TAIGA-IACT this is the first attempt to study the reconstruction of energy spectra using CNNs. Comparison of the results of each CNN showed that for the best reconstruction cases (GoogLeNet and DenseNet structures) the relative error reaches 22% in monoscopic observations, while in stereoscopic observations with two and three telescopes the error decreases to 19% and 12%, respectively. When the energy spectrum is reconstructed in mono mode, distortions are observed at the edges of the spectrum, which are significantly reduced in stereo mode.

Deep learning energy reconstruction was also compared with the traditional IACT image processing method based on the Hillas parameter method. Comparison was successful and it was demonstrated that deep learning reconstructed spectrum is in good agreement with reconstructed spectrum of the traditional method and model spectrum. However, the comparison of relative errors for each event showed that deep learning is slightly inferior in terms of the quality of reconstruction to the traditional method. We believe that reconstruction accuracy will be improved by more subtle settings of neural networks (for example, by changing the numerical parameters of the network and connections between layers). Also to improve results obtained it is planned to increase training datasets by increasing the number of training samples, and by expanding the energy range.

## Acknowledgments

The work was supported by the Russian Science Foundation, grant No. 22-21-00442. Also authors would like to thank the staff of the TAIGA collaboration for their help in the work and also express special gratitude to P.A. Volchugov for the provided data of traditional energy reconstruction method.